\documentclass[twocolumn, aps, superscriptaddress,showpacs, floatfix]{revtex4-1}
 \usepackage{graphicx}
 \usepackage{dcolumn}
 \usepackage{bm}
 \usepackage{ulem}
 \usepackage{enumitem}
\def\q{\bm{q}}
\def\p{\bm{p}}
\def\k{\bm{k}}
\def\v{\bm{v}}
\def\x{\bm{x}}

\begin{document}

\title{$\Upsilon(1S)$ transverse momentum spectra through dissociation and regeneration in heavy-ion collisions}
\author{Juhee Hong}
\affiliation{Department of Physics and Institute of Physics and Applied Physics, Yonsei University,
Seoul 03722, Korea}
\author{Su Houng Lee}
\affiliation{Department of Physics and Institute of Physics and Applied Physics, Yonsei University,
Seoul 03722, Korea}
\date{\today}

\begin{abstract}
We calculate the transition between a quarkonium state and an unbound 
heavy quark-antiquark pair through gluo-dissociation and inelastic parton 
scattering using a partonic picture that interpolates between the formal 
limits based on potential nonrelativistic QCD (pNRQCD) at different 
temperatures. 
While the thermal width increases with momentum and temperature, 
the quarkonium regeneration is affected by the heavy 
quark distribution function which depends on the diffusion constant. 
By solving the Boltzmann equation with the dissociation and regeneration terms, 
we investigate the medium modifications of quarkonium momentum spectra. 
Our numerical results indicate that the $\Upsilon(1S)$ $R_{AA}$ at high 
transverse momentum are influenced by the regeneration effects depending 
on the heavy quark diffusion. 
In this picture, the published CMS data that show an almost transverse 
momentum independence can be explained by the interplay between the 
suppression by dissociation and enhancement by regeneration at 
low and high transverse momenta, respectively. 
With the same input, we also calculate the transverse momentum dependence of 
the $\Upsilon(1S)$ $v_2$ and show that it lies within the limits of the 
available data. 
\end{abstract}

\maketitle

\section{Introduction}

As an important signal of the quark-gluon plasma formation \cite{satz}, 
quarkonium suppression has attracted much theoretical and experimental 
attention in heavy quarkonium physics. 
Although the yields of heavy quarkonia are reduced by the static color 
screening and dissociation by in-medium interactions, quarkonia states are 
known to survive above the phase transition temperature \cite{hatsuda} and 
can even be regenerated by recombination of a heavy quark and antiquark.
The quarkonium enhancement by regeneration not only increases the number 
of bound states but also affects their momentum spectra. 

Quarkonium dissociation has been investigated over the years 
\cite{Hansson:1987uk,kharzeev}, whereas its regeneration is less discussed 
and often ignored due to the low density of heavy quarks produced in 
heavy-ion collisions, especially for bottom quarks. 
The statistical hadronization and coalescence models have been used for 
$J/\Psi$ production \cite{stat,coal}.  
However, quarkonium momentum spectra are more sensitive to the regeneration 
mechanism than the total yields, and furthermore the centrality and 
energy-momentum dependences of the enhanced quarkonium distribution are 
difficult to predict within a statistical model or other models based on 
hadronic processes \cite{thews}. 
In addition to the production near the phase transition, we need to 
consider how the number and momentum distributions of the initially produced 
quarkonia and heavy quarks change by dissociation and regeneration during 
the evolution of the quark-gluon plasma. 
By treating both the thermal medium and quarkonium distributions dynamically, 
kinetic models with a rate or transport equation have been utilized to 
describe continuous regeneration of quarkonium through the quark-gluon 
plasma phase \cite{thews,pzhuang,rapp2010,rapp2011}.

The yield of $J/\Psi$ by recombination has been shown to be considerable, 
but the significance of bottomonium regeneration is problematic. 
This is so because the regeneration by recombining $b$ and $\bar{b}$ seems 
to be negligible as the density of bottom quarks and cross sections for 
open bottom are much smaller than those for charm quarks. 
On the other hand, if the transverse momentum dependent ratio of the 
bottomonium number to the number of $b$ quarks is much smaller than 
that of the corresponding values for the $c$ quarks, bottomonium regeneration 
might be relatively significant \cite{rapp2012,rapp2017}, especially in high 
energy heavy-ion collisions at $\sqrt{s_{NN}}=5.02$ TeV at the LHC.

In a partonic picture, quarkonium dissociation occurs through two scattering 
processes, gluo-dissociation ($\Upsilon+g\rightarrow b+\bar{b}$, where 
$\Upsilon$ is bottomonium) and inelastic parton scattering 
($\Upsilon+p\rightarrow b+\bar{b}+p$ with $p=g,q,\bar{q}$). 
The dipole interaction of color charges with gluon \cite{peskin} is used at 
leading order ($\sigma_{\rm LO}\sim g^2a_0^2$) for the first process and 
at next-to-leading order ($\sigma_{\rm NLO}\sim g^4a_0^2$) for the second 
process. 
By using the Bethe-Salpeter amplitude and hard thermal loop (HTL) 
perturbation theory, we have recently rederived the next-to-leading order 
dissociation cross section \cite{jhong}. 
In the relevant kinematical limit, our results reduce to the formal limits 
obtained by potential nonrelativistic QCD (pNRQCD) at which the thermal width 
is determined by the imaginary part of the singlet potential \cite{pnrqcd}.

The inverse reactions of gluo-dissociation and inelastic parton scattering 
contribute to the quarkonium regeneration. 
In order to describe the regeneration, the detailed balance condition 
or a coalescence model has been used in the kinetic approaches mentioned 
above. 
The detailed balance allows one to describe the quarkonium regeneration in 
terms of the dissociation process and the equilibrium distributions. 
It is valid only if quarkonium and heavy quarks are near equilibrium, 
which is not the case for the quark-gluon plasma phase.     
On the other hand, a coalescence model describes instantaneous regeneration 
regardless of specific reactions. 
It would be more appropriate to calculate the regeneration continually by 
parton-quarkonium interactions in the same approximation used for the 
dissociation. 
In this work, to describe the dissociation and regeneration mechanisms 
consistently, we use a Boltzmann transport equation with the collision terms 
which are obtained by the scattering amplitudes of the two processes, 
convoluting the momentum distributions of incoming and outgoing particles 
\cite{muller,muller2}. 
In this way, the regeneration term will depend on heavy quark distribution 
functions, and the momentum spectra of quarkonium will reflect the evolution 
of initial heavy quark spectra with modifications by a thermal medium.

We concentrate on bottomonium because our numerical approach based on the 
nonrelativistic heavy quark limit is more suitable for bottomonium than 
charmonium. 
Especially, $\Upsilon(1S)$ survives up to $\sim 600$ MeV \cite{review} so 
that it is crucial to analyze the evolution of the distribution over a wide 
temperature range within a consistent formalism. 
This can be accomplished within our approach, as our previously derived 
partonic formula consistently interpolates the constraints imposed by the 
pNRQCD formalism at high and low temperature limits.
The nuclear modification factor of $\Upsilon(1S)$ has been measured in PbPb 
collisions at $\sqrt{s_{NN}}=2.76, \, 5.02$ TeV by the CMS collaboration 
\cite{exp276,exp502}, and the elliptic flow at $\sqrt{s_{NN}}=5.02$ TeV 
by the ALICE and CMS collaborations \cite{v2exp,v2CMS}. 
The $\Upsilon(1S)$ $R_{AA}$ appears to be independent of the transverse 
momentum. 
The experimental data have been compared either with \cite{rapp2017} or 
without \cite{strickland} the regeneration of $\Upsilon(1S)$ within kinetic 
models, but the regeneration effects in Ref. \cite{rapp2017} are estimated 
by a coalescence model. 
The centrality and rapidity dependence of $\Upsilon(1S)$ $R_{AA}$ measured 
in pPb \cite{ppb1,ppb2,ppb3} and PbPb collisions have been reproduced in 
Ref. \cite{lansberg}.

Our goal is to understand how the dissociation and regeneration mechanisms 
influence on quarkonia momentum spectra, and eventually on the nuclear 
modification factor and the elliptic flow, using a consistent formalism that 
interpolates between the formal limits provided by the pNRQCD constraints. 
In Sec. \ref{Gamma}, 
we extend our previous work on the thermal width \cite{jhong} 
for quarkonium moving through the quark-gluon plasma. 
By taking the inverse reactions of gluo-dissociation and inelastic 
parton scattering, we calculate the regeneration term of the Boltzmann 
equation in Sec. \ref{C-reg}. 
The regeneration term involves heavy quark distribution functions 
which can be characterized by a Fokker-Planck equation with a diffusion 
constant. 
In Sec. \ref{result-fsol}, the Boltzmann equation with the dissociation 
and regeneration terms is solved for a Bjorken expansion at mid-rapidity. 
The numerical solutions for $\Upsilon(1S)$ are used to determine the nuclear 
modification factor in Sec. \ref{result-raa} and the elliptic flow 
in Appendix \ref{result-v2}. 
Sec. \ref{result-init} is devoted to the discussion about the uncertainties 
regarding initial conditions when comparing with experimental data.  
Finally, we summarize our results in Sec. \ref{summary}.

\section{Dissociation and regeneration of quarkonium moving in the quark-gluon 
plasma}
\label{Gamma-Creg}

The dynamic evolution of a quarkonium state can be described by 
dissociation, regeneration, and elastic scatterings. 
When the inverse distance between a heavy quark and its antiquark is larger 
than the temperature scale, elastic collisions are of higher order 
than dissociation and regeneration \cite{muller2}. 
Thus, neglecting this term, the Boltzmann equation for quarkonium is given by \cite{muller}  
\begin{eqnarray}
\label{Yeq}
&&\left(\frac{\partial}{\partial t}
+\v\cdot\frac{\partial }{\partial \x}\right)f_\Upsilon(t,\x,\q) 
=
-\Gamma_{\rm diss}^{\rm gluo+inel}(t,\x,\q) \, f_\Upsilon(t,\x,\q)
\nonumber\\
&& \qquad \qquad \qquad \qquad \qquad
+C_{\rm reg}^{\rm gluo+inel}[f_b,f_{\bar{b}}](t,\x,\q) \, ,
\end{eqnarray}
where $\v=\q/q^0$ is the quarkonium velocity, and the distribution 
functions $f_\Upsilon$ and $f_{b,\bar{b}}$ contain the fugacity factors of 
$\gamma_b^2$ and $\gamma_b$, respectively. 
The first term on the right hand side corresponds to the dissociation term 
with the thermal width $\Gamma_{\rm diss}$, and the second to the 
regeneration term which depends on the heavy quark and antiquark 
distribution functions.

\subsection{The thermal width}
\label{Gamma}

In Ref. \cite{jhong}, we have calculated the thermal width in the rest frame 
of quarkonium and integrated the phase space numerically for a 
phenomenological study. 
To extend this for quarkonium moving in the plasma, we proceed as follows: 
First, we calculate the scattering amplitudes (or dissociation cross sections) 
in the rest frame of quarkonium. 
Second, medium partons ($g,q,\bar{q}$) are considered to move with respect to 
quarkonium. 
Then their thermal distributions are 
$f(t,\x,\k)=1/[e^{\gamma(k^0-\k\cdot\v)/T}\pm 1]$, where 
$\gamma=1/\sqrt{1-\v^2}$ is the Lorentz factor. 
Third, the thermal width is obtained by convoluting the momentum distributions 
of moving partons with the scattering cross sections. 
Lastly, we divide the thermal width by the Lorentz factor to determine the 
width in the rest frame of the plasma \cite{lee-v}.

We have used the hard thermal loop (HTL) perturbation theory with an 
effective vertex derived from the Bethe-Salpeter amplitude to calculate 
the dissociation cross sections \cite{jhong}. 
The effective vertex describes the dipole interaction of color charge with 
gluon in the large $N_c$ limit (where the interaction between heavy quark 
and antiquark after dissociation can be neglected). 
For gluo-dissociation and inelastic parton scattering, the matrix elements 
have been obtained as 
\begin{eqnarray}
&&|\mathcal{M}|_{\rm gluo}^2
=\frac{8(N_c^2-1)}{N_c}g^2m^2m_\Upsilon k_0^2|\nabla\psi(\p)|^2 \, ,
\nonumber\\
&&|\mathcal{M}|_{\rm inel}^2
=16(N_c^2-1)g^4m^2m_\Upsilon|\nabla\psi(\p)|^2
\nonumber\\
&&\quad  \times
\frac{(\k_1-\k_2)^2k_{10}^2}
{[(\k_1-\k_2)^2+m_D^2]^2}
\left\{
\begin{tabular}{cc}
$[1+(\hat{\k}_1\cdot\hat{\k}_2)^2]$ & \quad ($g$) ,\\
$[1+\hat{\k}_1\cdot\hat{\k}_2]$ & \quad ($q,\bar{q}$) ,
\end{tabular}
\right.
\end{eqnarray}
for a Coulombic bound state, 
$|\nabla \psi(\p)|^2=2^{10}\pi a_0^{7}\p^2/[(a_0\p)^2+1]^6$, with the relative 
momentum $\p=(\p_1-\p_2)/2$. 
The Debye screening mass depends on temperature as 
$m_D^2=\frac{g^2T^2}{3}(N_c+\frac{N_f}{2})$. 
Following the above procedure, the thermal widths in the rest frame of the 
quark-gluon plasma are then given by 
\begin{eqnarray}
&&\Gamma_{\rm diss}^{\rm gluo}(t,\x,\q)
=\frac{1}{2d_\Upsilon \gamma q^0}
\int\frac{d^3\k}{(2\pi)^32k^0}
\nonumber\\
&&\qquad\qquad\qquad\times
\int\frac{d^3\p_1}{(2\pi)^32p_1^0}
\int\frac{d^3\p_2}{(2\pi)^32p_2^0}
|\mathcal{M}|^2_{\rm gluo}
\nonumber\\
&&\qquad\qquad\qquad \times
(2\pi)^4\delta^4(Q+K-P_1-P_2)
f(t,\x,\k) \, ,
\nonumber\\
&&\Gamma_{\rm diss}^{\rm inel}(t,\x,\q)
=\frac{1}{2d_\Upsilon \gamma q^0}\int\frac{d^3\k_1}{(2\pi)^32k_1^0}
\int\frac{d^3\k_2}{(2\pi)^32k_2^0}
\nonumber\\
&&\qquad\qquad\qquad\times
\int\frac{d^3\p_1}{(2\pi)^32p_1^0}
\int\frac{d^3\p_2}{(2\pi)^32p_2^0}
|\mathcal{M}|^2_{\rm inel}
\nonumber\\
&&\qquad\qquad\qquad\times
(2\pi)^4\delta^4(Q+K_1-K_2-P_1-P_2)
\nonumber\\
&&\qquad\qquad\qquad\times
f(t,\x,\k_1)
[1\pm f(t,\x,\k_2)] \, , 
\end{eqnarray}
where $Q, K, P_1,P_2$ denote the momenta of quarkonium, 
medium partons, heavy quark and heavy antiquark, respectively 
($\q,\p_1,\p_2\gg\k$). 
In inelastic parton scattering, the thermal distribution of an outgoing 
parton has the momentum $\k_2\simeq \k_1$ for small energy transfer, where 
$1,2$ refer to incoming and outgoing momenta, respectively.

\begin{figure}
\includegraphics[width=0.45\textwidth]{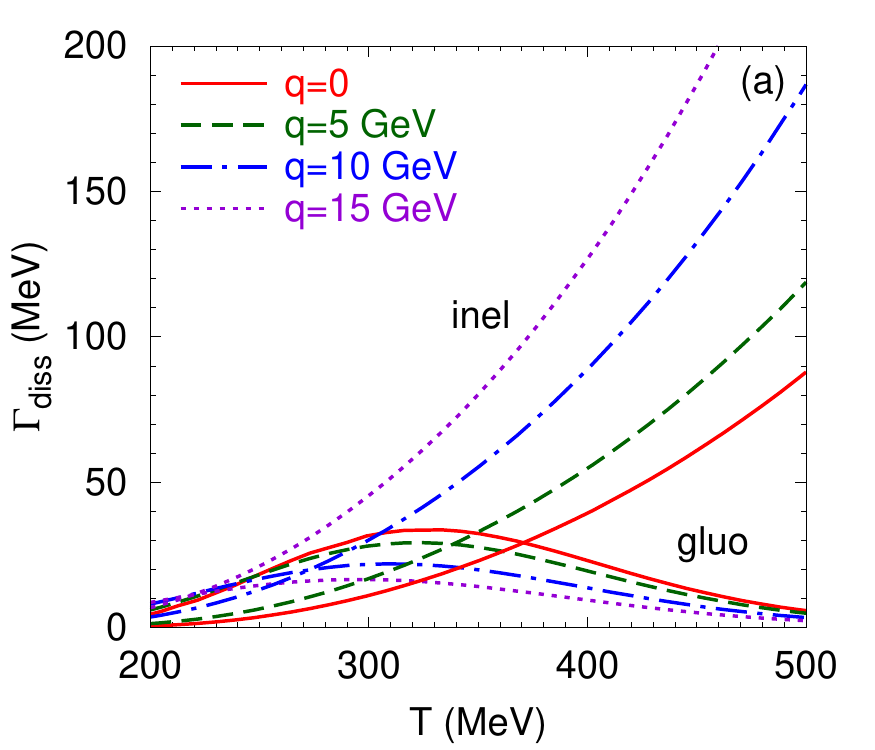}
\includegraphics[width=0.45\textwidth]{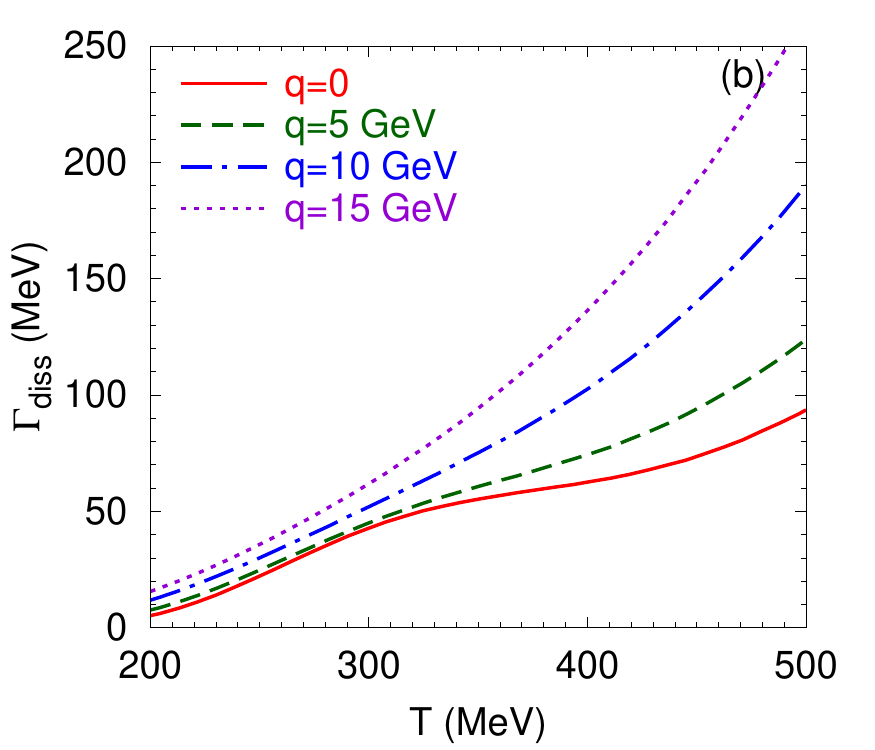}
\caption{The momentum dependence of the $\Upsilon(1S)$ thermal width as a 
function of the plasma temperature. 
(a) The comparison between gluo-dissociation and inelastic parton scattering. 
(b) The total thermal width as the sum of two processes.
}
\label{Gdiss}
\end{figure}

Figure \ref{Gdiss} shows the numerical results for the $\Upsilon(1S)$ thermal 
widths which depend on the momentum. 
We have used the same values of the parameters as Ref. \cite{jhong}: 
$\alpha_s=0.4$, $a_0=0.14$ fm, $m=4.8$ GeV, and the binding energy $E$ 
estimated in lattice QCD \cite{lattice}. 
The thermal width of inelastic parton scattering increases with momentum 
but the width of gluo-dissociation decreases. 
This behavior comes from the Lorentz factor and the thermal distributions 
of partons involved in the width. 
Specifically, the factor of $f(t,\x,\k_1)[1\pm f(t,\x,\k_2)]/\gamma$ increases 
with the quarkonium momentum especially for high $\k_1$ where inelastic parton 
scattering is dominant. 
Furthermore, the inelastic scattering cross section increases with $\k_1$. 
For low $\k$, gluo-dissociation becomes effective and the factor of 
$f(t,\x,\k)/\gamma$ decreases as the $\Upsilon(1S)$ momentum grows. 
These behaviors toward momentum are reflected in the thermal width as seen in 
Fig. \ref{Gdiss} (a). 
For low temperature at which the binding energy is larger than the Debye mass, 
gluo-dissociation is effective especially for quarkonium at rest. 
On the other hand, inelastic parton scattering is dominant over 
gluo-dissociation at high temperature where $m_D\gg E$, as in pNRQCD 
\cite{pnrqcd}.  
Since the momentum dependence of inelastic parton scattering is stronger than 
that of gluo-dissociation, the sum of two contributions grows with the 
momentum (see Fig. \ref{Gdiss} (b)). 
Our numerical results of the thermal width qualitatively agree with those of 
Ref. \cite{lee-v}.

\subsection{The regeneration term}
\label{C-reg}

The regeneration terms in the Boltzmann equation are obtained by the inverse 
reactions of gluo-dissociation and inelastic parton scattering, 
\begin{eqnarray}
\label{Cregeq}
&&C_{\rm reg}^{\rm gluo}(t,\x,\q)
=\frac{1}{2d_\Upsilon \gamma q^0}\int\frac{d^3\k}
{(2\pi)^32k^0}
\int\frac{d^3\p_1}{(2\pi)^32p_1^0}
\nonumber\\
&& \qquad\quad
\times
\int\frac{d^3\p_2}{(2\pi)^32p_2^0}
|\mathcal{M}|^2_{\rm gluo}
(2\pi)^4\delta^4(Q+K-P_1-P_2)
\nonumber\\
&&\qquad \quad \times
f_b(t,\x,\p_1)f_{\bar{b}}(t,\x,\p_2)
[1+f(t,\x,\k)] \, ,
\nonumber\\
&&C_{\rm reg}^{\rm inel}(t,\x,\q)
=\frac{1}{2 d_\Upsilon \gamma q^0}
\int\frac{d^3\k_1}{(2\pi)^32k_1^0}
\int\frac{d^3\k_2}{(2\pi)^32k_2^0}
\nonumber\\
&&\quad \times 
\int\frac{d^3\p_1}{(2\pi)^32p_1^0}
\int\frac{d^3\p_2}{(2\pi)^32p_2^0}
|\mathcal{M}|^2_{\rm inel}
\nonumber\\
&&\quad \times
(2\pi)^4\delta^4(Q+K_1-K_2-P_1-P_2)
\nonumber\\
&&\quad \times 
f_b(t,\x,\p_1)
f_{\bar{b}}(t,\x,\p_2)f(t,\x,\k_2)
[1\pm f(t,\x,\k_1)] \, ,
\end{eqnarray}
where the phase space integrations are numerically performed as done in the 
thermal width calculation \cite{jhong}. 
The momenta of partons and heavy quarks are approximated as $\k_1\simeq \k_2$ 
and $\p_1=-\p_2\simeq\p$ in the rest frame of quarkonium, respectively.

Through the regeneration term, the momentum spectrum of 
bottomonium reflects the initial $b,\bar{b}$ spectra which themselves evolve by 
interacting with partons in a thermal medium.  
Bottom quarks are produced with a power-law transverse momentum spectrum 
which is harder than the thermal one. 
The approach to the thermal equilibrium is slower than that of light quarks 
by a factor $\sim m/T$ \cite{moore-teaney}, and  thus 
they are not likely to be thermalized in the quark-gluon plasma phase. 
In principle, the $b$ distribution function should be obtained by solving 
a Boltzmann equation, but we use a simpler Fokker-Planck equation which 
characterizes the time evolution of $b$ quarks.

To determine the $b$ distribution function, we consider elastic scatterings, 
$p+b\rightarrow p+b$. 
In the limit of small energy transfer, the $t$-channel gluon exchange is 
dominant, and the collision term at leading-log in $T/m_D$ can be 
approximated as a Fokker-Planck operator \cite{svet,hees-rapp,moore-teaney}. 
For a uniform plasma, we then have  
\begin{equation}
\frac{\partial f_{b}(t,\p)}{\partial t}=
\left[\frac{\partial}{\partial \p}\cdot\eta(t)\p
+\frac{1}{2}\frac{\partial^2 }{\partial \p^2}\kappa(t)\right]f_{b}(t,\p) \, ,
\end{equation}
where $\eta(t)$ is the drag coefficient and $\kappa(t)$ is the 
momentum diffusion constant. 
By ignoring the momentum dependences of drag and diffusion, 
(when the heavy quark velocity vanishes) $\kappa(t)=2mT(t)\eta(t)$. 
The momentum diffusion constant is related to the diffusion constant in space, 
$D(t)=2T^2/\kappa(t)$.  
Then $f_b(t,\p)$ can be obtained from the initial condition $f_b(t_0,\p_0)$ by 
convolving the Green's function as follows \cite{moore-teaney,hees-rapp}: 
\begin{equation}
f_b(t,\p)=\int d^3\p_0 \, G(t,\p;t_0,\p_0)f_b(t_0,\p_0) \, , 
\end{equation}
where  
\begin{equation}
G(t,\p;t_0,\p_0)=\frac{1}{[4\pi K(t)]^{3/2}}\exp\left[
-\frac{(\p-\p_0e^{-H(t)})^2}{4K(t)}\right] \, ,
\end{equation}
with $H(t)=\int_{t_0}^t dt' \, \eta(t')$ and 
$K(t)=\frac{1}{2}e^{-2H(t)}\int_{t_0}^t dt'\, \kappa(t') \, 
e^{2H(t')}$.

All the $b$ quarks are expected to be produced in initial hard collisions. 
The fugacity factor $\gamma_b(t)$ is determined by requiring 
that the total number of $b$ quarks is constant throughout the quark-gluon 
plasma phase,  
\begin{equation}
N_{b}(t)=d_{b}\gamma_b(t)V(t)\int \frac{d^3\p}{(2\pi)^3} 
f_{b}(t,\p) \, ,
\end{equation}
where $d_{b}$ is the degeneracy factor of $b$ quarks.
Because the production cross section of a hidden bottom state is much smaller 
than that of $b\bar{b}$ pairs ($\frac{\sigma_{pp\rightarrow\Upsilon}}
{\sigma_{pp\rightarrow b\bar{b}}}\sim 10^{-3}$) \cite{rapp2017}, we ignore 
the $\Upsilon$ contribution to $N_b(t)$. 
At the LHC and RHIC, at most a few $b,\bar{b}$ pairs are produced so we 
present the numerical results for the case $N_b(t)=1$ in the following.

\begin{figure}
\includegraphics[width=0.45\textwidth]{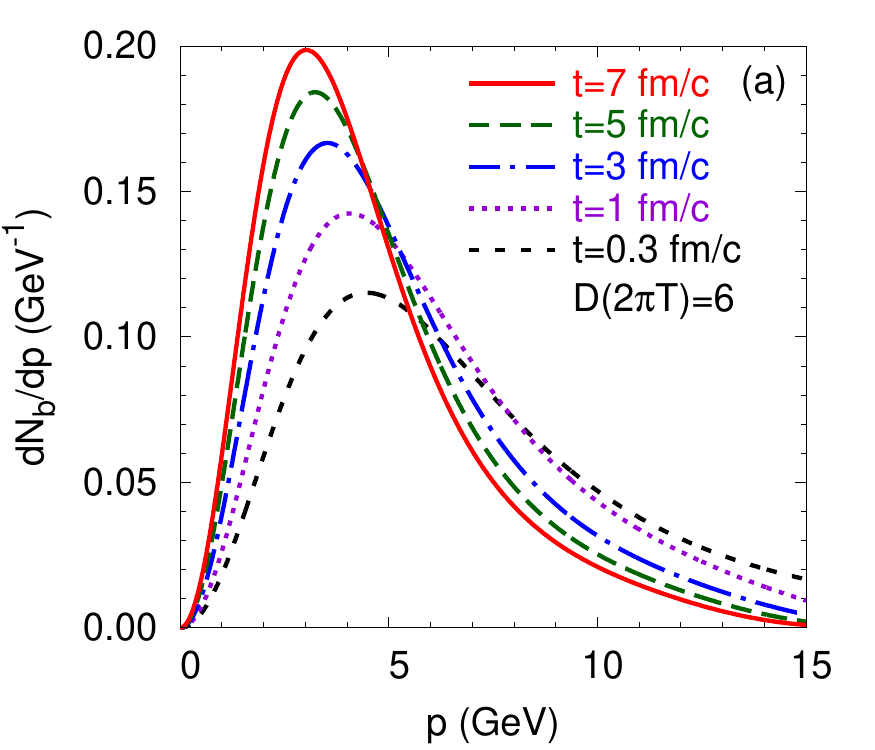}
\includegraphics[width=0.45\textwidth]{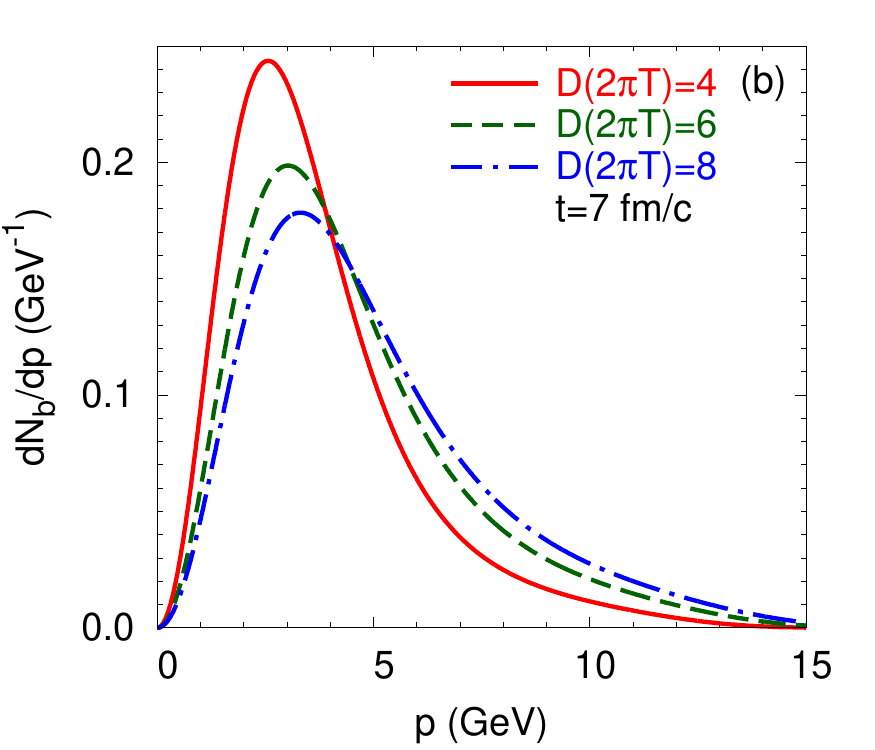}
\caption{
(a) The time evolution of the $b$ quark distribution in momentum space. 
(b) The dependence on the diffusion constant.
}
\label{fb}
\end{figure}

At early times of heavy-ion collisions, the transverse dimension of a 
central collision system is so large that the dynamics is dominated 
by a longitudinal motion. 
We suppose a thermal medium to be in a local equilibrium at time $t_0$ with 
temperature $T_0$.  
For a Bjorken expansion \cite{bjorken}, the time dependences of drag, 
temperature and volume are given by $\eta(t)=\eta_0(t_0/t)^{2/3}$, 
$T(t)=T_0(t_0/t)^{1/3}$, and $V(t)=V_0\,t/t_0$, respectively, 
where we set $T_0=550$ MeV and $V_0=60 \, \mbox{fm}^3$ at 
$t_0=0.3$ fm/c. 
Since there are neither much data nor theoretical calculations for $b$ quarks, 
for an initial distribution we use the $B$ meson differential 
cross section measured in pp collisions \cite{exp-fb} as a starting point 
(the uncertainties related to the initial conditions are discussed in 
Sec. \ref{result-init}). 
Figure \ref{fb} (a) shows the time evolution of the $b$ quark momentum 
distribution.
As the diffusion constant decreases, the distribution function becomes 
more localized in momentum space, as displayed in Fig. \ref{fb} (b). 
The thermal distribution is much softer than that with $D(2\pi T)=4$.

\begin{figure}
\includegraphics[width=0.45\textwidth]{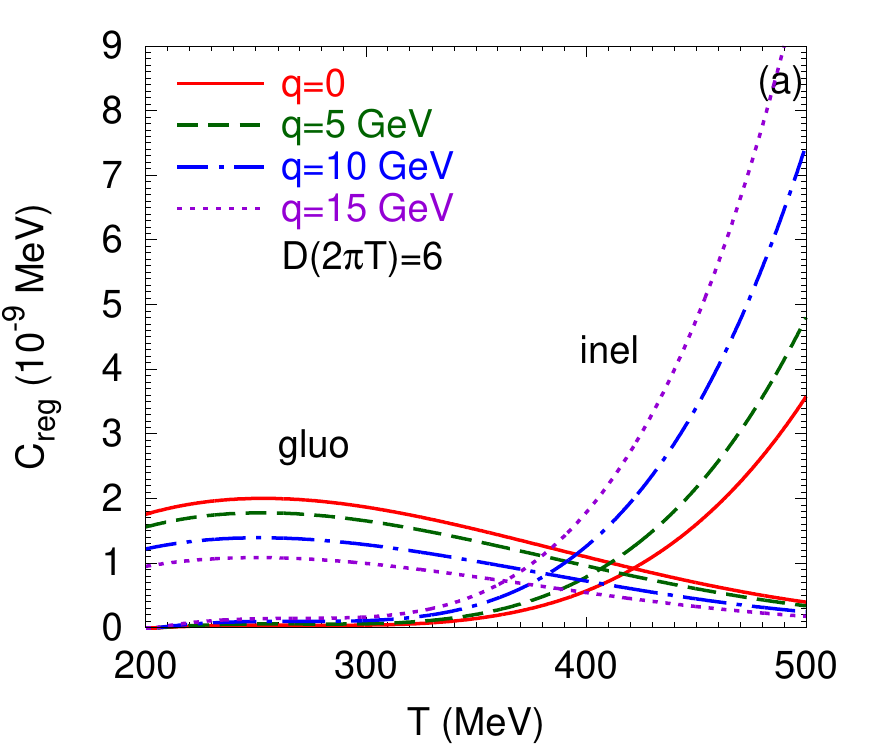}
\includegraphics[width=0.45\textwidth]{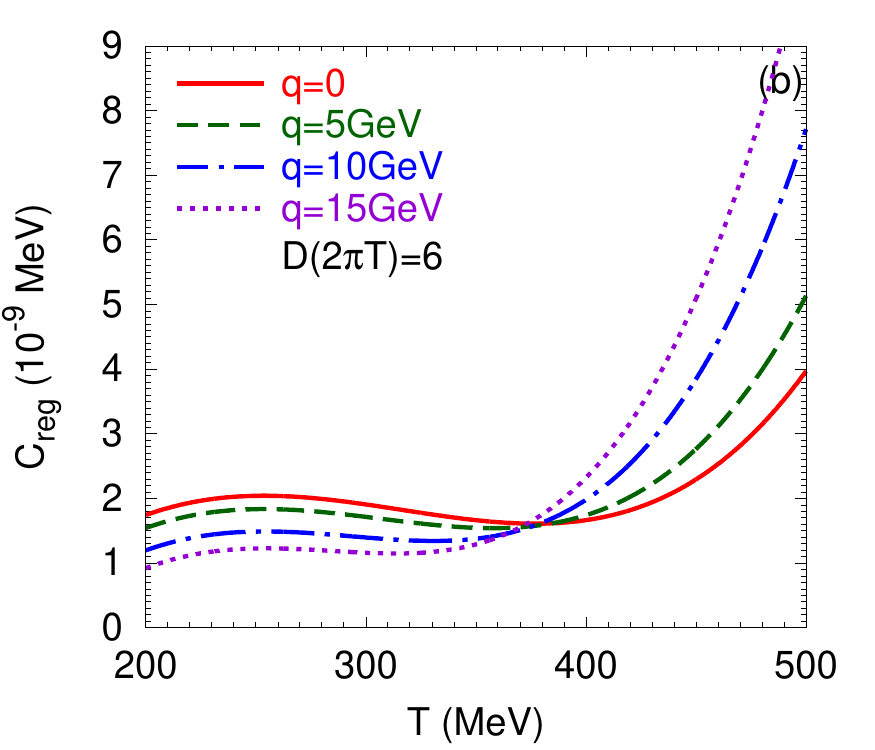}
\caption{
The momentum dependence of the $\Upsilon(1S)$ regeneration term in the 
Boltzmann equation, Eq. (\ref{Yeq}). 
(a) The comparison of the inverse gluo-dissociation with inelastic parton 
scattering. 
(b) The sum of two contributions.
}
\label{Creg}
\end{figure}

\begin{figure}
\includegraphics[width=0.45\textwidth]{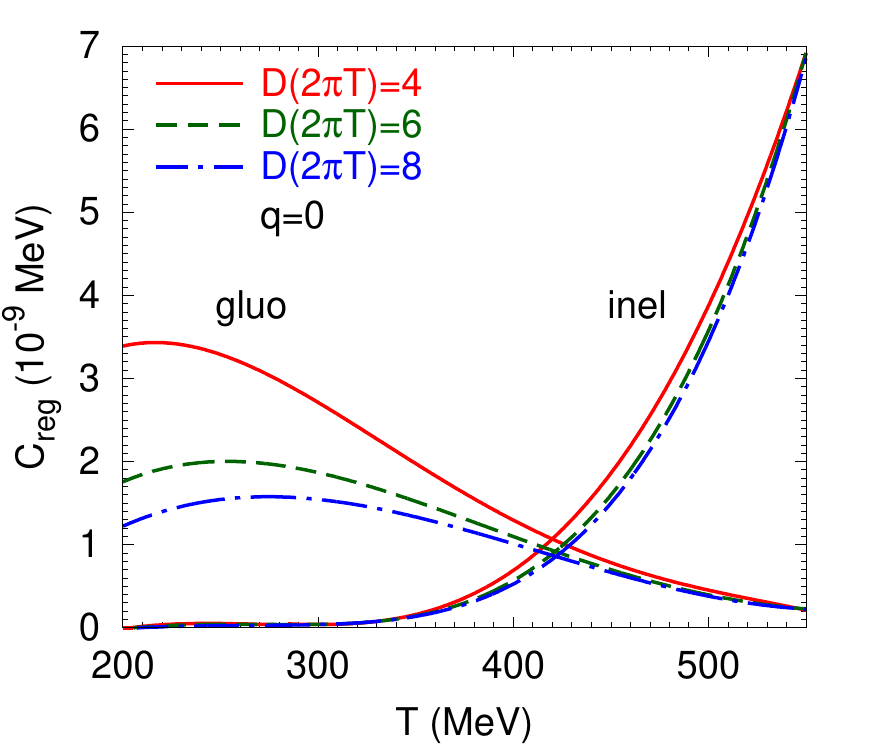}
\caption{ 
The regeneration term depending on the $b$ quark diffusion constant. 
}
\label{Creg3}
\end{figure}

By using the $b,\bar{b}$ momentum distributions computed above, the 
regeneration term of Eq. (\ref{Cregeq}) can be obtained numerically 
(see Fig. \ref{Creg}). 
As with the thermal width, the regeneration by inelastic parton scattering 
is dominant at high temperature whereas that by the inverse gluo-dissociation 
is efficient near the phase transition. 
As the momentum grows, inelastic parton scattering 
(inverse gluo-dissociation) regenerates more (less) $\Upsilon(1S)$. 
The regeneration term exhibits a similar momentum dependence to the 
thermal width case, because the factor $f(t,\x,\k_2)[1\pm f(t,\x,\k_1)]/\gamma$ or 
$[1+f(t,\x,\k)]/\gamma$ acts in the same way as discussed in Sec. \ref{Gamma}. 
However, the regeneration effects by the inverse gluo-dissociation can be 
rather strong at low temperature, especially when the heavy quark diffusion 
is small. 
Figure \ref{Creg3} shows the $D$ dependence of the regeneration term by two 
mechanisms. 
A less diffuse $f_b(t,\p)$ makes the $\Upsilon(1S)$ regeneration more 
probable, and 
the regeneration through the inverse gluo-dissociation is more influenced by 
the diffusion constant. 
The heavy quark diffusion has scanty effect on the regeneration by inelastic 
parton scattering, because the process is important only at high temperature 
where the distribution of $b,\bar{b}$ is close to the initial input.

Strong drag can lead to the enhanced production of a bound state, 
whereas fast diffusion should decrease quarkonium yields \cite{svet}. 
The drag and diffusion constants are related to the equilibration rate of 
heavy quarks and affect the regeneration process.   
For smaller $\eta$ and larger $D$, the equilibration is slower and the 
regeneration effects are weaker.   
Because the numerically calculated distribution of $b,\bar{b}$ are harder 
than the thermal distribution, the regeneration term presented here is 
smaller than that calculated with the thermal distributions.

\section{Medium modifications of quarkonium transverse momentum spectra}
\label{result}

In the previous section, we have obtained the thermal width and the 
regeneration term which depend only on time and momentum. 
Now, we apply it in the central rapidity region with the Lorentz invariance 
under a longitudinal boost. 
Since the right hand side of Eq. (\ref{Yeq}) is independent of space, 
taking the average over $\x_T$ yields the transverse momentum spectrum, 
$f_\Upsilon(t,\q_T)$:  
\begin{eqnarray}
\label{Yeq2}
\frac{\partial f_\Upsilon(t,\q_T)}{\partial t} &=& 
-\Gamma_{\rm diss}^{\rm gluo+inel}(t,\q_T) \, f_\Upsilon(t,\q_T)
\nonumber\\
&&\qquad\qquad
+C_{\rm reg}^{\rm gluo+inel}[f_b,f_{\bar{b}}](t,\q_T) \, .
\end{eqnarray}
We are interested in the medium modifications of $R_{AA}(q_T)$ and the elliptic 
flow $v_2(q_T)$ (see Appendix \ref{result-v2}) which are induced by the 
momentum-dependent dissociation and regeneration. 
Because the quarkonium regeneration takes place only below the dissociation 
temperature $T_{\rm diss}$ at which the thermal width becomes 
comparable to or exceeds the reduced binding energy, we consider the 
quark-gluon plasma phase within the temperature range 
$T_c \lesssim T \lesssim T_{\rm diss}$.

\subsection{The $\Upsilon(1S)$ distribution function}
\label{result-fsol}

The quarkonium distribution function is obtained by solving Eq. (\ref{Yeq2}) 
numerically for each momentum. 
As with $b$ quarks, the initial number and momentum distribution of 
$\Upsilon(1S)$ are not well known. 
For an initial condition, we use the differential cross section for 
$\Upsilon(1S)$ as a function of its transverse momentum and per unit rapidity 
measured in pp collisions \cite{exp502}, normalized by 
$\frac{N_\Upsilon(t_0)}{N_b(t_0)+N_{\bar{b}}(t_0)}\approx 
\frac{\sigma_{pp\rightarrow\Upsilon}}{\sigma_{pp\rightarrow b\bar{b}}}
= 1.76\times10^{-3}$ \cite{rapp2017}.

\begin{figure}
\includegraphics[width=0.45\textwidth]{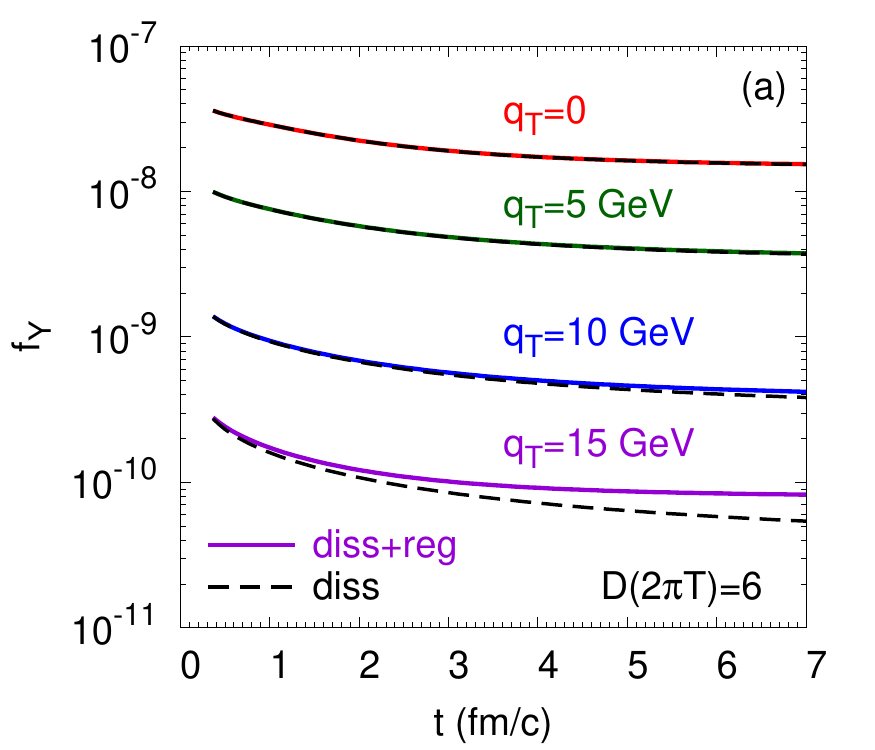}
\includegraphics[width=0.45\textwidth]{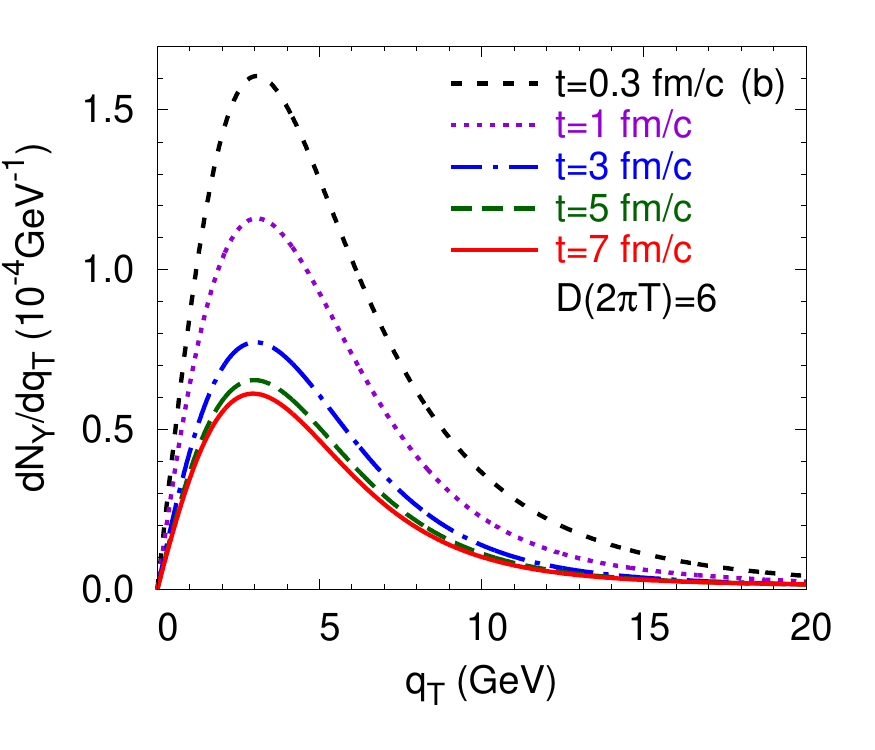}
\caption{
(a) The numerical solution of Eq. (\ref{Yeq2}). 
The regeneration effects are included in the solid lines, in comparison with 
the dashed lines which are obtained by the dissociation term only. 
(b) The time evolution of the $\Upsilon(1S)$ momentum distribution.  
}
\label{fsol}
\end{figure}

As a first step, we can ignore the regeneration term because bottomonium 
regeneration depends on the densities of $b,\bar{b}$ quarks which are fairly 
small. 
Then the distribution function has an exponential form,
\begin{equation}
\label{sol-diss} 
f_\Upsilon^{\rm diss}(t,\q_T)=f_\Upsilon^{\rm diss}(t_0,\q_T)\, 
e^{-\int_{t_0}^t dt'\, \Gamma_{\rm diss}^{\rm gluo+inel}(t',\q_T)} \, , 
\end{equation} 
as denoted by the dashed lines in Fig. \ref{fsol} (a). 
If we include the regeneration term, the suppression is slightly reduced 
(see the solid lines). 
At higher momentum where the initial number is smaller, 
the regeneration effects are more prominent. 
From Figs. \ref{Gdiss} (b), \ref{Creg} (b), and \ref{fsol} (a), we note 
that $C_{\rm reg}/f_\Upsilon\sim 10$ MeV at $q_T=15$ GeV but  
$\Gamma_{\rm diss}\sim 100$ MeV. 
Thus, using these numbers in the first and second terms of Eq. (\ref{Yeq2}), 
one expects that the regeneration effects are at most $\sim 10 \%$ compared to 
the dissociation contribution.  
In Fig. \ref{fsol} (b), we exhibit the time evolution of the $\Upsilon(1S)$ 
momentum distribution. 
The accumulated dissociation effect increases with time but its change 
decreases, so that the numerical solution freezes around $t\approx 7$ fm/c at 
the end of the quark-gluon plasma phase.  
Since the bottomonium dissociation is dominant over regeneration, the 
number of $\Upsilon(1S)$ is reduced to approximately $40\%$.

\subsection{The nuclear modification factor}
\label{result-raa}

The nuclear modification factor can be estimated by the ratio of the final 
spectrum of $\Upsilon$ to the initial one, 
\begin{equation}
R_{AA}(q_T)=\frac{\frac{dN_\Upsilon}{q_Tdq_T}\big\vert_{t=t_f}}
{\frac{dN_\Upsilon}{q_Tdq_T}\big\vert_{t=t_0}} \, ,
\end{equation}
where $t_f\approx 7$ fm/c is the time when the dissociation and regeneration 
mechanisms stop working at $T_f\approx T_c$.  
After the phase transition, hadronic effects are expected to be insignificant 
because of small reaction cross sections.

\begin{figure}
\includegraphics[width=0.45\textwidth]{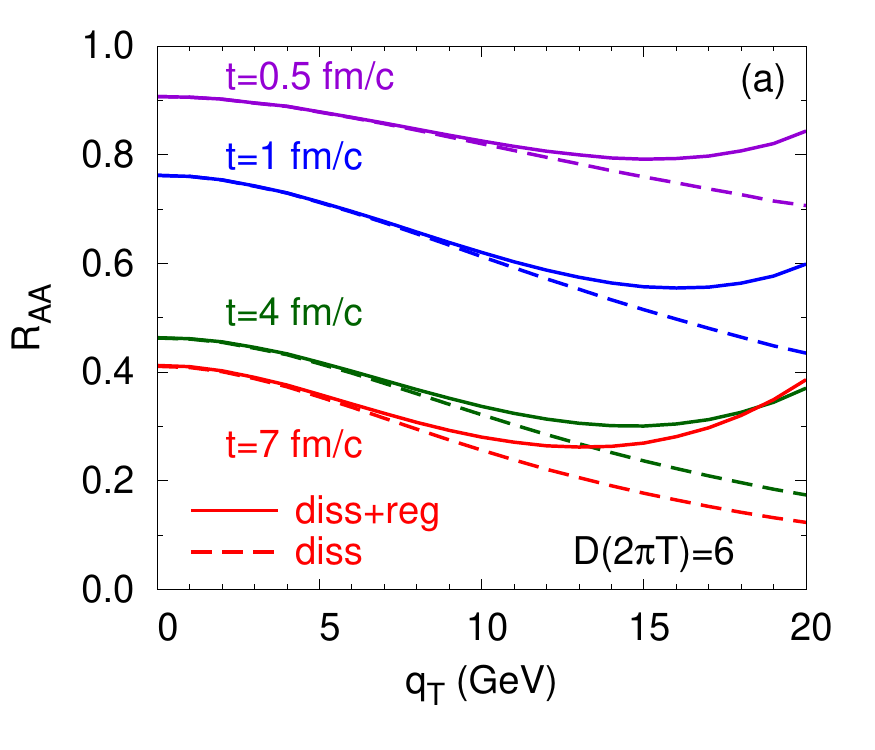}
\includegraphics[width=0.45\textwidth]{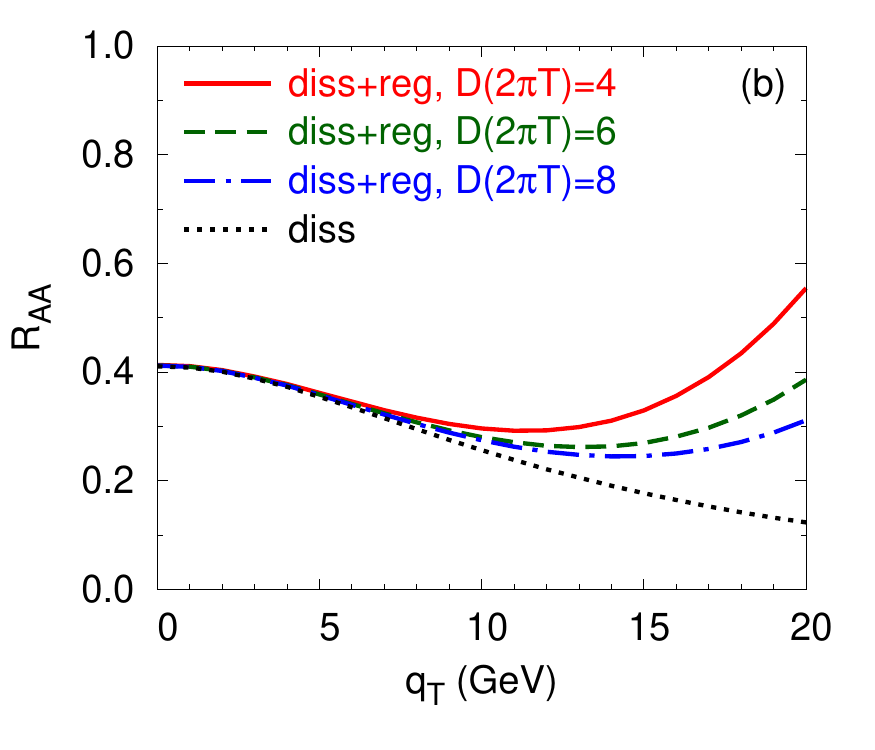}
\caption{
(a) The regeneration effects on the nuclear modification factor for 
$\Upsilon(1S)$. 
(b) The dependence on the $b$ quark diffusion constant. 
}
\label{raa}
\end{figure}

Figure \ref{raa} (a) shows the time evolution of the medium modifications. 
In the absence of regeneration, $R_{AA}$ corresponds to the 
exponentially decaying factor of Eq. (\ref{sol-diss}) and is denoted by the 
dashed lines. 
As a collision system evolves, the temperature decreases and 
so does the thermal width (see Fig. \ref{Gdiss} (b)). 
Thus, the suppression increases ($R_{AA}$ decreases) with time but eventually 
freezes at $t\approx 7$ fm/c. 
Furthermore, the thermal width grows with the transverse momentum, and 
the $R_{AA}$ decreases as the momentum increases. 
However, as seen in the solid lines of Fig. \ref{raa} (a), at high $q_T$ 
the regeneration effects are substantial because the dissociation term comes 
with a small number of $\Upsilon(1S)$.

The diffusion constant of $b$ quark also affects the nuclear modification 
factor. 
From Fig. \ref{raa} (b), we see that the suppression at high $q_T$ changes for 
the different diffusion constant as in the case for the regeneration term. 
As the number of $\Upsilon(1S)$ is reduced to $40\%$ in Fig. \ref{fsol} 
(b), $R_{AA}\simeq 0.4$ at low momentum which is independent of the 
diffusion constant.  
On the other hand, $R_{AA}$ at high momentum depends considerably on the 
diffusion of $b,\bar{b}$. 
The $D$ dependence of $R_{AA}$ is agreeable with the findings in Ref. 
\cite{chen} which is based on the Langevin equation and Wigner function.

\subsection{The dependence on initial conditions}
\label{result-init}

There are significant uncertainties in the initial stage of heavy-ion 
collisions. 
Although we have assumed a locally equilibrated medium at $t_0$ with 
$T_0\lesssim T_{\rm diss}$, there is a pre-equilibrium stage where 
quarkonia formation is in progress and the dissociation mechanism begins to 
work before regeneration \cite{rapp2010}. 
Since the pre-equilibrium effects can lead to different initial conditions, 
we discuss the possible consequences in this section.

For initial distributions of $b,\bar{b}$ and $\Upsilon(1S)$, we have used 
the differential cross section measured in pp collisions, 
fit to the following form:    
\begin{equation}
f(t_0,p_T) \propto \frac{1}{\big[\big(\frac{\p_T}{\Lambda}\big)^2+1\big]^{\alpha}} \, .
\end{equation}
For $f_b(t_0,p_T)$ and $f_\Upsilon(t_0,q_T)$, $\alpha=2.85, \,2.44$ and 
$\Lambda=6.07,\,6.05$ GeV, respectively.   
The fitted spectra are questionable as the current experimental data have large 
uncertainties especially at high momentum \cite{exp502,exp-fb}. 
In order to compare our calculations with the measurements, we need to 
study the dependence of our results on the initial distributions of 
$b,\bar{b}$ and $\Upsilon(1S)$.

\begin{figure}
\includegraphics[width=0.45\textwidth]{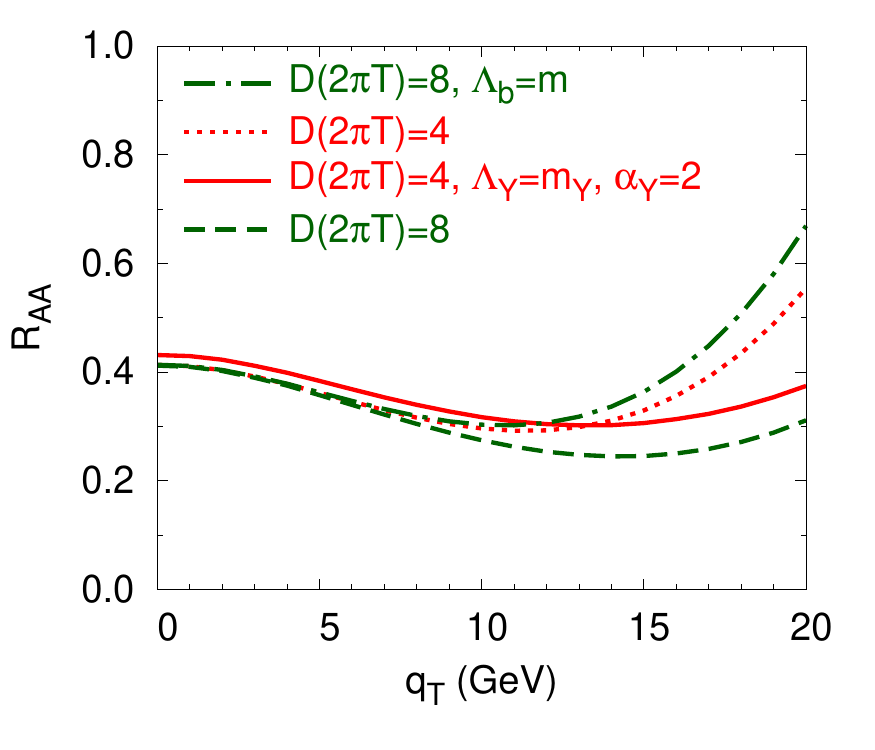}
\caption{
The $R_{AA}$ depending on the initial distributions of $f_b(t_0,p_T)$ and 
$f_\Upsilon(t_0,q_T)$. 
The default $\Lambda$ and $\alpha$ values given in the text are used for 
those not specified in the figure.
}
\label{Dy}
\end{figure}

Figure \ref{Dy} shows the $R_{AA}$ dependence on the initial distributions 
of $b,\bar{b}$ and $\Upsilon(1S)$. 
If we use a harder quarkonium spectrum with larger $\Lambda=m_\Upsilon$ and 
smaller $\alpha=2$ \cite{strickland}, 
the nuclear modification factor decreases at high momentum (compare the red 
dotted line with the solid line). 
On the other hand, when $b,\bar{b}$ spectra become softer with smaller 
$\Lambda=m$ (within the experimental uncertainties), 
$R_{AA}$ increases as shown in the green dashed-dotted line compared to the 
dashed line. 
As with a smaller diffusion constant, a softer initial spectrum leads to 
a larger nuclear modification factor at high momentum. 
We notice that the $D$ dependence of $R_{AA}$ shown in Fig. \ref{raa} (b) 
might not be satisfied when different initial distributions for $b,\bar{b}$ 
and $\Upsilon(1S)$ are used for each $D$ value.

\begin{figure}
\includegraphics[width=0.45\textwidth]{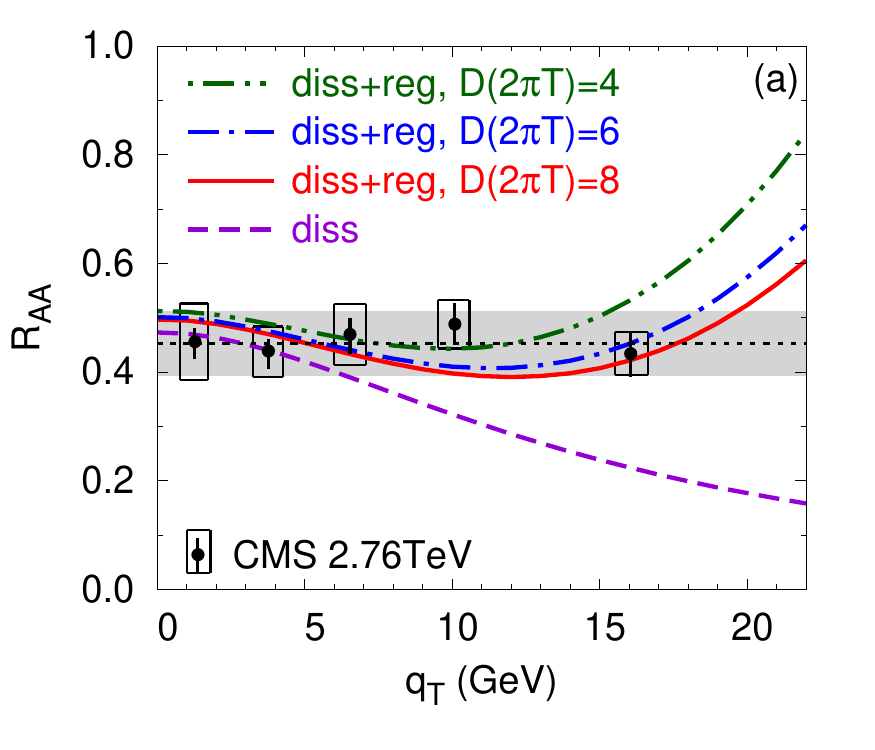}
\includegraphics[width=0.45\textwidth]{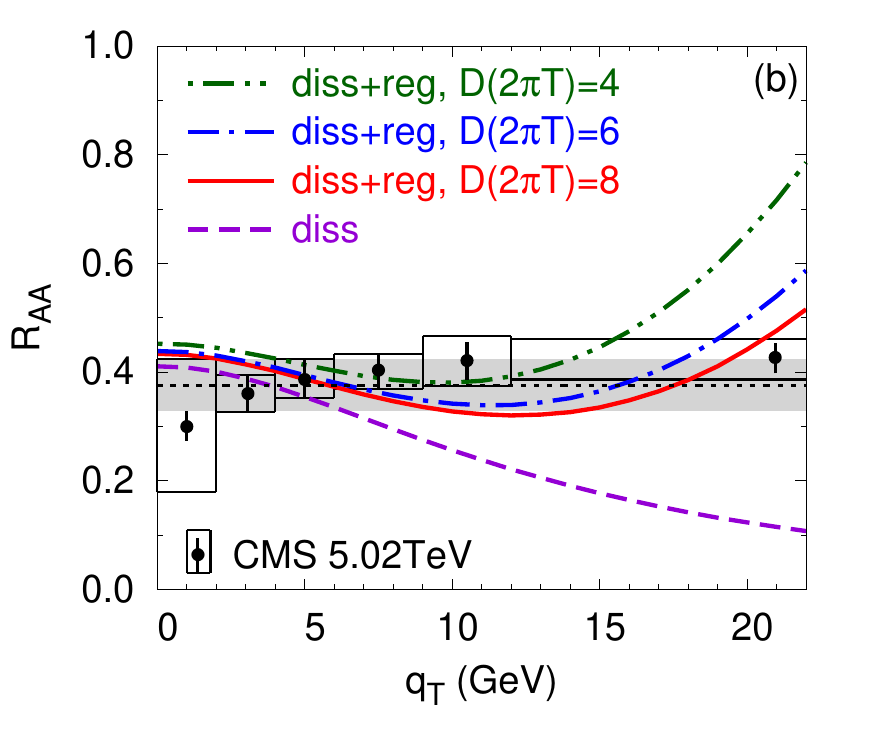}
\caption{
The calculated $\Upsilon(1S)$ $R_{AA}$ is compared with the CMS data measured 
in PbPb collisions at $\sqrt{s_{NN}}=2.76, \, 5.02$ TeV \cite{exp276,exp502}.  
$f_b(t_0,p_T)$ with $\alpha=2.85$, $\Lambda=m$ and $f_\Upsilon(t_0,q_T)$ 
with $\alpha=2$, $\Lambda=m_\Upsilon$ have been used, and the initial 
temperature $T_0=525, \, 550$ MeV have been assumed for 
$\sqrt{s_{NN}}=2.76, \, 5.02$ TeV, respectively. 
The dotted lines with the gray band indicate the centrality-integrated 
$R_{AA}$ including the statistical and systematic uncertainties. 
}
\label{compare}
\end{figure}

The $\Upsilon(1S)$ $R_{AA}$ has been measured in PbPb collisions at 
$\sqrt{s_{NN}}=2.76, \, 5.02$ TeV by the CMS Collaboration 
\cite{exp276,exp502}. 
The centrality-integrated 
$R_{AA}= 0.453\pm 0.014(\mbox{stat})\pm 0.046(\mbox{syst})$, 
\, $0.376\pm 0.013(\mbox{stat})\pm 0.035(\mbox{syst})$ at 
$\sqrt{s_{NN}}=2.76, \, 5.02$ TeV, respectively, and 
$R_{AA}$ appears to be independent of the transverse momentum. 
Figure \ref{compare} compares our calculations with the experimental data. 
The initial distributions $f_b(t_0,p_T)$ with $\alpha=2.85$, $\Lambda=m$ and 
$f_\Upsilon(t_0,q_T)$ with $\alpha=2$, $\Lambda=m_\Upsilon$ have been used. 
Using the same initial distributions, the $R_{AA}$ with smaller diffusion is 
larger as discussed before. 
The initial temperature $T_0=525, \, 550$ MeV has been assumed for 
$\sqrt{s_{NN}}=2.76, \, 5.02$ TeV, respectively. 
When the initial temperature is larger at higher energy collisions, the phase 
transition takes place later and the $R_{AA}$ is smaller at low momentum 
where dissociation is dominant (compare Fig. \ref{compare} (a) with (b)).  
The experimental data are comparable to our results with 
$D(2\pi T)\approx 5-8$. 
We note that both gluo-dissociation and inelastic parton scattering need 
to be taken into account to describe the nuclear modification factor 
$\approx 0.4$. 
Since the medium suppression exponentially increases with momentum 
due to dissociation (as denoted by the violet dashed line), the regeneration 
effects might account for the seemingly momentum independence of the measured 
data.

To estimate the feed-down effects, we can apply our numerical approach to 
the excited states of bottomonium (for which the gluo-dissociation can be 
more important than inelastic parton scattering because $T_{\rm diss}<300$ 
MeV \cite{review}). 
The direct production of $\Upsilon(1S)$ is approximately $67\%$ and the 
remainder is mostly from $1P$ and $2S$ states 
\cite{rapp2017,strickland}. 
However, the binding energies of the excited states are smaller at least by 
a factor of $3-5$ than the ground state energy \cite{lattice}, 
so their thermal widths are much larger than the $\Upsilon(1S)$ width. 
As a result, a major part of the inclusive $R_{AA}$ is expected to 
come from the direct $\Upsilon(1S)$ suppression. 
For instance, the $R_{AA}^{\Upsilon(2S)}\sim 0.1$ \cite{exp276,exp502} is 
roughly four times smaller than $R_{AA}^{\Upsilon(1S)}$. 
If the other excited states have nuclear modification factors of the same 
order as $R_{AA}^{\Upsilon(2S)}$, 
then the feed-down effects amount to $\sim 10\%$ of the direct 
$\Upsilon(1S)$ contribution.   
In principle, the feed-down can reduce the inclusive $R_{AA}$, especially at 
high $q_T$ in Fig. \ref{compare}, but the spectrum would be still 
enhanced by regeneration in comparison to the suppressed one by 
dissociation only.

We do not expect that our results are considerably affected by varying 
other parameters involved in our calculations: 
(1) $T_0$ and $t_0$ can alter the lifetime of 
the quark-gluon plasma and the shapes of the $R_{AA}$ curves. 
(2) With the larger $N_b$ and the smaller $N_\Upsilon/N_b$, the regeneration 
effects become more important. 
These parameters might change the numerical results quantitatively, but the 
tendency of dissociation and regeneration to the quarkonium momentum 
remains the same.

\section{Summary}
\label{summary}

We have discussed the dissociation and regeneration effects on 
the quarkonium momentum distributions. 
By taking into account the gluo-dissociation, inelastic parton scattering, 
and their inverse reactions through a partonic cross section formula that 
interpolates the formal limits at different temperature region, we have 
calculated the thermal width and the regeneration 
term of the Boltzmann equation for quarkonium moving in the quark-gluon 
plasma. 
For a Bjorken expansion geometry, the nuclear modification factor of $\Upsilon(1S)$ 
has been determined and compared with the experimental data. 
Due to the dominant dissociation, the $R_{AA}$ is exponentially 
decaying but the regeneration effect tends to  reduce the medium suppression at high transverse momentum. 
The quarkonium regeneration depends on the heavy quark distribution which has 
been determined by a Fokker-Planck equation with a heavy quark diffusion 
constant. 
For smaller diffusion, the heavy quark distribution is more localized and 
the regeneration effects are more significant. 
With the heavy quark diffusion constant $D(2\pi T)\approx 5-8$ and 
a rather hard initial distribution of $\Upsilon(1S)$, our 
numerical results seem to agree with the experimental measurements of the 
$R_{AA}$. 
Our analysis implies that the bottomonium spectra at high momentum are 
influenced by the regeneration effects, whereas those at low momentum 
are controlled by dissociation.

The $\Upsilon(1S)$ spectrum has been described by the Boltzmann equation with 
the dissociation and regeneration terms, given initial conditions. 
The two collision terms have been obtained consistently based on the 
Bethe-Salpeter amplitude and hard thermal loop (HTL) resummation.  
As our thermal widths reduce to the effective field theory results in 
the relevant kinematical limit, which should be more applicable for 
$b$ quarks than $c$ quarks, the uncertainties involved in our calculations 
will be mainly from the initial conditions and the unknown heavy 
quark diffusion constant.

The upcoming experimental data and the lattice QCD computations 
on the diffusion constant can be useful to reduce the uncertainties 
in our approach.  
Furthermore, we need to include the nuclear geometric evolution and 
other mechanisms such as feed-down and quarkonium diffusion (and 
even the quantum decoherence effects caused by the noise correlations 
\cite{akamatsu1,akamatsu2}) for a 
more sophisticated phenomenological study.

\section*{Acknowledgments}

This work is supported by the National Research Foundation of Korea
(NRF) grant funded by the Korea government (MSIT) (No. 2018R1C1B6008119), and  by Samsung Science and Technology Foundation under Project Number SSTF-BA1901-04.

\bigskip

\appendix

\section{The elliptic flow}
\label{result-v2}

In the main text, we have assumed that the (transverse) reaction plane is 
isotropic. 
Here, we consider the azimuthal angular anisotropy 
$v_2^b$ of $b,\bar{b}$ distributions 
\begin{equation}
\frac{dN_b}{d\phi}\propto
1+2v_2^b(p_T)\cos(2\phi) \, ,
\end{equation}
and study its contribution to the $\Upsilon(1S)$ $v_2$ \cite{muller}. 
For the momentum dependent anisotropy of $b$, we use the $b$ quark $v_2(p_T)$ 
calculated in Ref. \cite{v2b}.  
Then, the elliptic flow induced by the regeneration term is given by 
\begin{equation}
v_2(q_T)=\frac{\int d\phi \frac{dN_\Upsilon}{q_Tdq_Td\phi}
\cos(2\phi)}{\int d\phi \frac{dN_\Upsilon}{q_Tdq_Td\phi}}\bigg\vert_{t=t_f} \, .
\end{equation}

\begin{figure}
\includegraphics[width=0.45\textwidth]{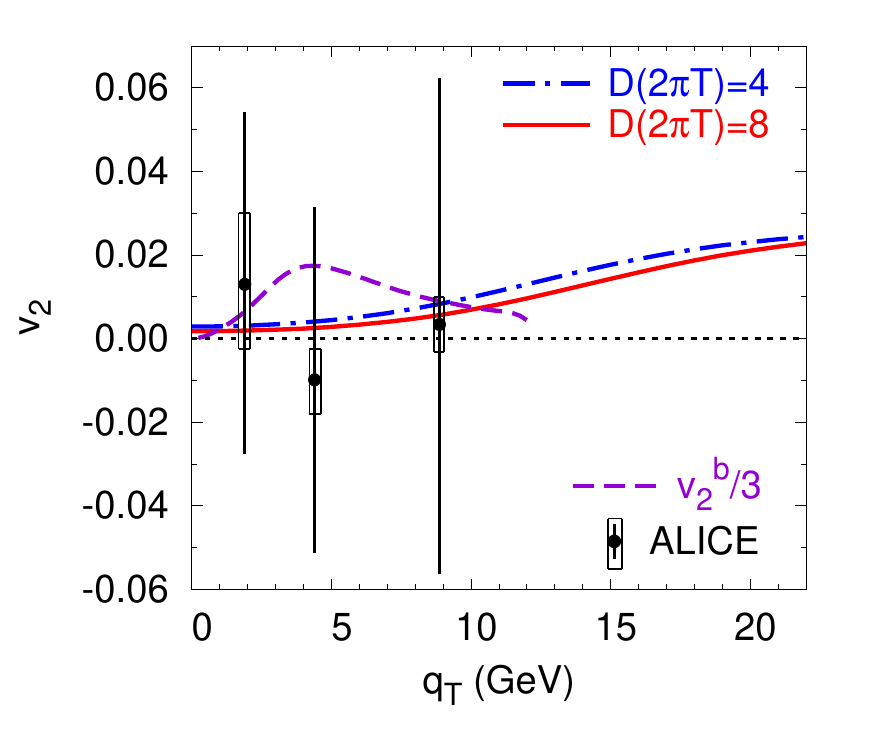}
\caption{
The regeneration effects on the elliptic flow anisotropy for $\Upsilon(1S)$, 
comparing with the ALICE data at forward rapidity measured in PbPb collisions 
at $\sqrt{s_{NN}}=5.02$ TeV \cite{v2exp}. 
The same initial conditions as Fig. \ref{compare} have been used. 
The violet dashed line indicates the azimuthal angular anisotropy of 
$b,\bar{b}$ quarks (divided by $3$) \cite{v2b} used in the calculations. 
}
\label{v2}
\end{figure}

Figure \ref{v2} shows the regeneration contribution to the elliptic flow of 
$\Upsilon(1S)$, supposing the $v_2^b(p_T)/3$ is given by the violet dashed 
line. 
When $v_2^b(p_T)$ increases with momentum up to $p_T\sim m$ and then decreases, 
$v_2(q_T)$ grows with $q_T$ and seems to level out at high $q_T$. 
As the heavy quark diffusion decreases, the regeneration effects become 
stronger so $v_2$ increases. 
The calculated $v_2$ is comparable with the ALICE data 
at forward rapidity measured in PbPb collisions at $\sqrt{s_{NN}}=5.02$ TeV 
\cite{v2exp}. 
In addition to the regeneration, there are other contributions from the 
dissociation term, such as the path-length difference, on the elliptic flow, 
but  such effects will be left for a future investigation.

\end{document}